\documentclass[aps,onecolumn,preprintnumbers,showpacs,showkeys,nofootinbib,
superscriptaddress  
]{revtex4}
\usepackage[american]{babel}
\usepackage{epsfig}
\usepackage{amssymb,amsmath,amsfonts,amsthm,graphicx,psfrag}
\usepackage{hyperref}
\newcommand{\beq}{\begin{eqnarray}}
\newcommand{\eeq}{\end{eqnarray}}

\begin{document}
\preprint{}

\title{
Temperature dependence of bulk viscosity within lattice simulation of $SU(3)$--gluodynamics
}

\author{N.~Yu.~Astrakhantsev}
\email[]{nikita.astrakhantsev@itep.ru}
\affiliation{ Moscow Institute of Physics and Technology, Dolgoprudny, 141700 Russia }
\affiliation{ Institute for Theoretical and Experimental Physics, Moscow, 117218 Russia }

\author{V.~V.~Braguta}
\email[]{braguta@itep.ru}
\affiliation{ Moscow Institute of Physics and Technology, Dolgoprudny, 141700 Russia }
\affiliation{ Institute for Theoretical and Experimental Physics, Moscow, 117218 Russia } 
\affiliation{ Bogoliubov Laboratory of Theoretical Physics, Joint Institute for Nuclear Research, Dubna, 141980 Russia } 
\affiliation{Far Eastern Federal University, School of Biomedicine, 690950 Vladivostok, Russia }

\author{A.~Yu.~Kotov}
\email[]{kotov@itep.ru}
\affiliation{ Moscow Institute of Physics and Technology, Dolgoprudny, 141700 Russia }
\affiliation{ Institute for Theoretical and Experimental Physics, Moscow, 117218 Russia }
\affiliation{ Bogoliubov Laboratory of Theoretical Physics, Joint Institute for Nuclear Research, Dubna, 141980 Russia }

\begin{abstract}
In this paper the temperature dependence of the $SU(3)$--gluodynamics bulk viscosity is studied within lattice simulations. 
To carry out this study we measure the correlation function of the trace of the
energy-momentum tensor for a set of temperatures within the range $T/T_c \in (0.9, 1.5)$.
To extract the bulk viscosity from the correlation function we apply the
Backus-Gilbert method and the Tikhonov regularization method. We show that the ratio $\zeta/s$ is 
small in the region $T/T_c \geqslant 1.1-1.2$ and in the vicinity of the transition $T/T_c \leqslant 1.1-1.2$ it quickly rises.
Our results are in agreement with previous lattice studies and in a reasonable agreement with other phenomenological approaches. Obtained values of the bulk viscosity are significantly larger than perturbative results, what confirms that QGP is a strongly correlated system.
\end{abstract}



\maketitle

\section*{\large Introduction}
Hydrodynamics is believed to describe time evolution of quark-gluon plasma (QGP) created in heavy ion collision experiments (such experiments are carried out at RHIC and LHC and planned in the future at FAIR and NICA). The basic object in hydrodynamics is 
the energy-momentum tensor built as an expansion in gradients \cite{Son:2008zz,Romatschke:2009im}. The leading order of this expansion describes an ideal fluid. The next-to-leading order includes dissipation and can be parameterized by 
two coefficients: shear and bulk viscosities. Trying to describe the particle yield in heavy ion collisions
one can determine the shear and bulk viscosities of QGP \cite{Ryu:2017qzn}. In particular, 
typical value of the shear viscosity to the entropy density ratio extracted from the hydrodynamic studies is $\eta/s=(1-2.5)\times 1/4\pi$ \cite{Song:2012ua}.

It is therefore important to calculate these observables based on our theoretical knowledge of the system. Since QGP is a strongly correlated system, one of the main ways to carry out the first principle study of its properties is the lattice simulation of QCD. 
Despite considerable success in lattice study of QGP, 
lattice calculations of the shear and bulk viscosities still remain challenging problems requiring huge statistics. 
For this reason nowadays it is not feasible to calculate viscosities in QCD with dynamical quarks. 
In the following we are going to address viscosities in gluodynamics.

Lattice calculations of the gluodynamics shear viscosity were carried out in \cite{Karsch:1986cq, Nakamura:2004sy, Meyer:2007ic, Meyer:2009jp, Mages:2015rea, Astrakhantsev:2015jta, Astrakhantsev:2017nrs, Pasztor:2018yae}. They are in agreement with each other and with the experimental data \cite{Song:2012ua} within the uncertainties. 
The lattice results are also close to $\eta/s=1/4\pi$ obtained within the $N = 4$ supersymmetric Yang-Mills theory at strong coupling~\cite{Policastro:2001yc}. 
For many years there was a disagreement between lattice results and the perturbative calculations \cite{Arnold:2000dr,Arnold:2003zc}. 
However, recent calculations in the next-to-leading order \cite{Ghiglieri:2018dib} give much smaller values of the shear viscosity, which are consistent with the lattice data. It is also worth to mention the paper \cite{Christiansen:2014ypa}, where the authors determined the value of the shear viscosity using the diagrammatic representation (their results are also very small and close to the lattice and the experimental data).

Another important transport coefficient is the QGP bulk viscosity $\zeta$. There are 
only few rather old papers devoted to lattice calculation of the bulk viscosity \cite{Nakamura:2004sy,Meyer:2007dy}. 
Taking into account the rapid improvement of supercomputers and theoretical developments it is reasonable to 
conduct an up-to-date study of the bulk viscosity. 

Let us now consider what is known about the bulk viscosity. 
One can expect that at very large temperature the results of perturbative calculation are applicable. It gives a very small value $\zeta/s \sim 0.02 \alpha_s^2$ for $N_f=0$ \cite{Arnold:2006fz}.
Unfortunately, this result cannot be applied for temperatures $\sim$ few $\times$ the critical temperature $T_c$, which is of interest for heavy ion collision experiments.  The bulk viscosity in this region was studied in \cite{Kharzeev:2007wb,Karsch:2007jc}. The authors applied the low energy theorems of QCD and 
derived the formula relating the spectral function of the energy-momentum tensor trace correlator  
to the energy density and pressure of hot matter. Using a physically motivated ansatz of the spectral function 
the authors found $\zeta$. At the critical temperature the bulk viscosity has a peak with the height $\zeta/s \sim 1$. 
For temperatures $T>T_c$ the bulk viscosity quickly drops becoming very small $\zeta/s<0.1$ already for $T>1.1 T_c$. There are a lot of 
phenomenological studies of the bulk viscosity \cite{J.P.:2018sts,Saha:2017xjq,Samanta:2017ohm, Singha:2017jmq, Ozvenchuk:2012kh, Marty:2013ita, Berrehrah:2016vzw} confirming the existence of peak at the critical temperature.

In these paper we study the temperature dependence of the bulk viscosity in $SU(3)$--gluodynamics 
within lattice simulation. We calculate the correlation functions of the energy-momentum 
tensor trace for the set of temperatures. To extract the spectral function and the bulk viscosity from 
the correlator we use two model-independent estimation techniques: the Backus-Gilbert method \cite{Backus:1,Backus:2} and the  
Tikhonov regularization approach \cite{Tikhonov:1963}.

This paper is organized as follows. In the next section we describe the details of the lattice 
measurements of the correlation functions under study. The results of this measurement are 
presented in Section III. In the last section we discuss our results and draw the conclusion. 

\section*{\large Details of the calculation}

Bulk viscosity is related to the Euclidean correlation function of the trace of the energy-momentum tensor:
\begin{equation}\begin{split}
	C(\tau) = T^{-5} \int d^3{\textbf x}\langle \theta(0) \theta(\tau,{\textbf x})\rangle,
\end{split}
 \label{correlator}
\end{equation}
where $\theta = \frac{\beta(g)}{2 g} F_{\mu \nu}^a F_{\mu \nu}^a$, $\beta(g)$ is the $\beta$-function of gluodynamics, $T$ is the temperature.
The correlator (\ref{correlator}) can be expressed in terms of the spectral function $\rho(\omega)$ via the integral equation
\begin{equation}\begin{split}
	C(\tau)= T^{-5} \int\limits_{0}^{\infty}\rho(\omega)\frac{\cosh \omega(\beta/2 - \tau)}{\sinh \omega \beta/2 }d\omega.
\label{spectr_corr}
\end{split}
\end{equation}

The spectral function contains valuable information about medium properties. In particular, one can
find the bulk viscosity if the spectral function is known \cite{Jeon:1995zm} 
\begin{equation}\begin{split}
	\zeta=\frac{\pi}{9}\lim\limits_{\omega\to0} \frac{d\rho(\omega)}{d\omega}.
\end{split}\end{equation}

Lattice calculation of bulk viscosity can be divided into two parts. First, one measures the correlation function $C(\tau)$ with sufficient accuracy. Second, one determines the spectral function $\rho(\omega)$ from $C(\tau)$. Although the first step is very complicated, sufficient accuracy can be achieved due to the multi-level algorithm \cite{Meyer:2002cd}. The second step is the well known ill-posed problem which is difficult to solve. 

Important properties of the spectral function are positivity $\rho(\omega)\geqslant 0,~\omega>0$ and oddness $\rho(-\omega)=-\rho(\omega)$. It is 
also important that at the leading order approximation in the strong coupling constant the spectral function can be written as
\begin{equation}
\rho_{LO} (\omega) = d_A\left(\frac{11 \alpha_s N_c}{3 (4 \pi)^2} \right)^2\frac{\omega^4}{\tanh (\omega  / 4T)},
\label{rho_tree_level}
\end{equation}
where $d_A = N_c^2 - 1 = 8$ for the $SU(3)$--gluodynamics. We expect that due to the asymptotic freedom the leading order expression (\ref{rho_tree_level}) is a good approximation for the spectral function at large frequency. To account for the discretization errors in temporal direction instead of (\ref{rho_tree_level}) we are going to use the tree level lattice expression $\rho_{lat}(\omega)$ calculated within the approximation: $L_t$ is fixed and $L_s\to \infty$ \cite{Meyer:2009vj}. The resulting expression for $\rho_{lat}(\omega)$ is cumbersome, for this reason we do not show it here.

At small frequencies the spectral function is in the hydrodynamic regime. The first order hydrodynamic
behavior for the spectral function reads
\begin{equation}
\rho_{h}(\omega)=\frac 9 {\pi} \zeta \omega 
\label{rhoh}
\end{equation}

In numerical simulation we use the Wilson gauge action for the $SU(3)$--gluodynamics. For $F_{\mu\nu}$ the clover discretization scheme is used. Similar to the shear viscosity the bulk viscosity is presented as the viscosity-to-entropy-density ratio $\zeta/s$. For homogeneous systems the entropy density $s$ can be expressed as $s = \frac{\epsilon+p}T$, where $\epsilon$ is the energy density and $p$ is the pressure. These thermodynamic quantities were measured with the method described in \cite{Engels:1999tk}. 

The energy-momentum tensor in the continuum theory is a set of the Noether currents which are related to the Lorentz invariance of the action. In the lattice formulation of field theory continuum rotational invariance does not exist and the renormalization of energy-momentum tensor is required. For the trace of the energy-momentum tensor the renormalization is multiplicative. The renormalization factors depend on the discretization scheme~\cite{Meyer:2007tm}. For the plaquette-based discretization scheme the renormalization is defined by the $\beta$-function~\cite{Meyer:2007tm}. Using the renormalization factors for the plaquette-based discretization of $\theta$, one can easily find the renormalization factors for the clover discretization by fitting the expectation value of the trace anomaly: $Z^{(plaq)}\langle\theta^{(plaq)}\rangle=Z^{(clov)}\langle\theta^{(clov)}\rangle$.
 
\section*{\large Numerical results}
\subsection*{Correlation functions and their properties}

We measured the correlation functions $C(\tau)$ on the lattice $16\times32^3$ with the parameters listed in Table~\ref{table:params}.  

\begin{table}[h]
\begin{center}
	\begin{tabular}{ c | c | c | c | c | c | c | c | c | c | c | c | c | c }
 	 $T / T_c$ & $0.90$ & $0.925$ & $0.950$ & $0.975$ & $1.00$ & $1.05$ & $1.10$ & $1.15$ & $1.19$ & $1.275$ & $1.35$ & $1.425$ & $1.50$ \\ \hline
  	 $\beta$ & 6.491 & 6.512 & 6.532 & 6.552 & 6.575 & 6.61 & 6.647 & 6.682 & 6.712 & 6.765 & 6.811 & 6.855 & 6.897 \\
	\end{tabular}
\end{center}
	\caption{The set of parameters used in the calculation of $C(\tau)$.}
    \label{table:params}
\end{table}

\begin{figure}[h!]
	\begin{center}
		\includegraphics[scale = 0.5]{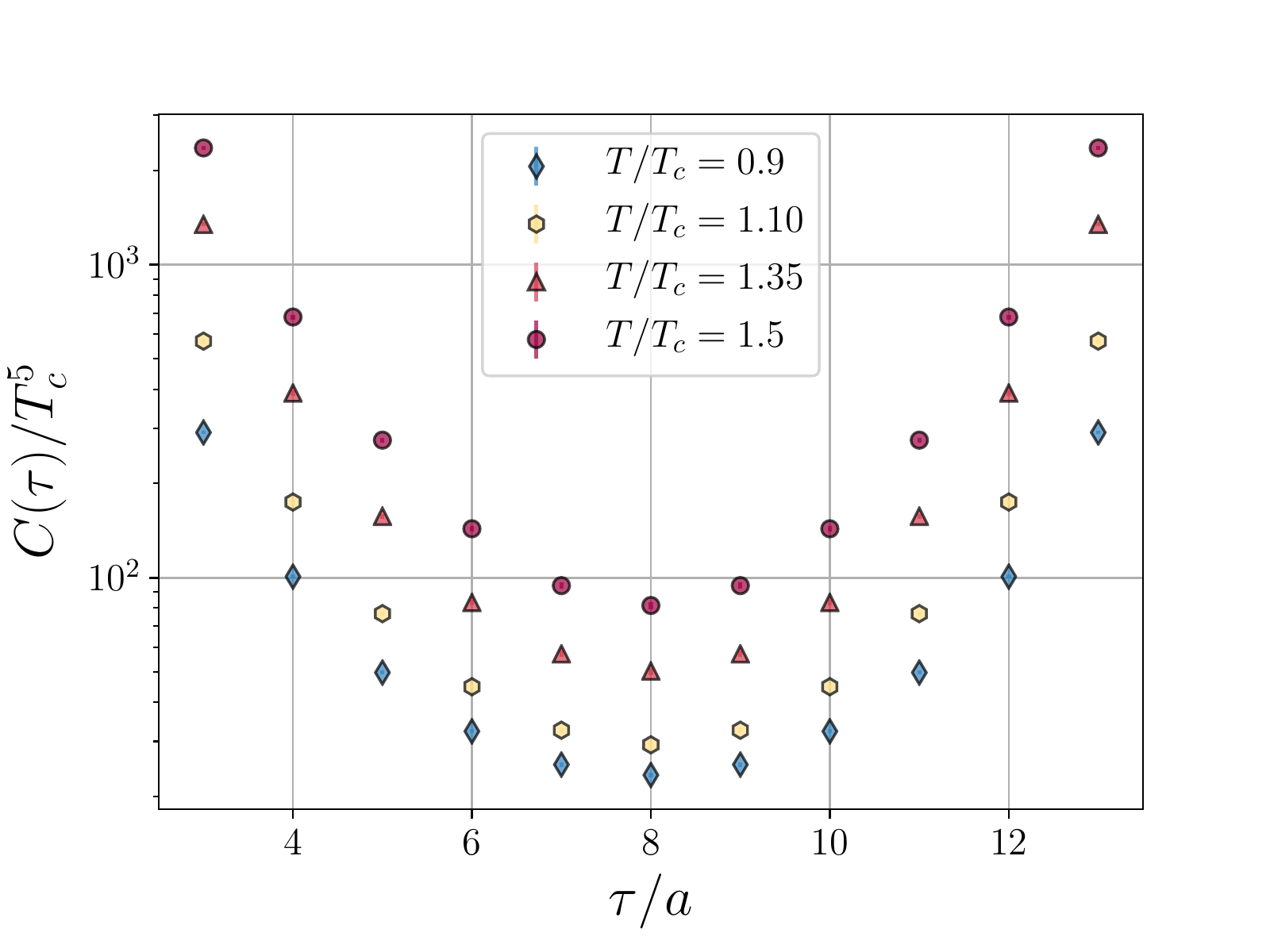}
		\caption{The correlation function $C(\tau)$ as a function of Euclidean time $\tau$ for the temperatures $T / T_c = 0.90,\,1.10,\,1.35,\,1.5$}
		\label{fig:correlation_f}
	\end{center}
\end{figure}

The two-level algorithm allowed us to reach relative errors of 2-3\% at the middle point $\tau T = 1 / 2$ for all the temperatures. For other values of Euclidean time $\tau$ the relative errors are even smaller. In Fig.~\ref{fig:correlation_f} we show correlation functions (\ref{correlator}) for the temperatures $T / T_c = 0.90,\,1.10,\,1.35,\,1.5$.

In order to estimate the finite volume effects, we measured the correlation functions (\ref{correlator}) on the larger lattice $16 \times 48^3$ for the temperatures $T/T_c=0.9, 0.975, 1.0, 1.05, 1.5$. We found that 
for the temperatures $T/T_c=0.9, 0.975$ the deviations of the correlators measured on two 
lattices are less than $2 \sigma$. For the temperatures $T/T_c = 1.05, 1.5$ the deviations are less than $\sigma$. 
For the temperature $T/T_c = 1.0 $ the deviation of the correlators 
in the middle point vicinity is as large as $4 \sigma$. 
The deviations for other points are smaller. Thus, we 
expect finite volume effects to be small for all the temperatures except $T=T_c$. Finite volume effects for the temperature $T=T_c$ might be important. 

\subsection*{Calculation of the bulk viscosity using the Backus-Gilbert method}

In this section we determine the ratio $\zeta/s$ using the 
Backus-Gilbert (BG) method \cite{Backus:1, Backus:2}. This is a model independent approach estimating the spectral function\footnote{ In \cite{Astrakhantsev:2017nrs} the BG method was used to study the shear viscosity and in \cite{Boyda:2016emg} it was applied to calculate the conductivity of graphene.}. Instead of $\rho(\omega)$ one reconstructs the estimator $\bar \rho(\bar \omega)$ expressed as 
\beq
\bar \rho(\bar \omega) = f(\bar \omega) \int_0^{\infty} d \omega \delta (\bar \omega, \omega) \frac {\rho(\omega)} {f(\omega)}, 
\label{barf}
\eeq
where the $f(x)$ is an arbitrary function and the $\delta(\bar \omega, \omega)$ is called the resolution function. This 
function has a peak around $\bar \omega$ and normalized as 
$\int_0^{\infty} d \omega \delta(\bar \omega, \omega)=1$. The BG resolution function is taken in the form
\beq
\delta(\bar \omega, \omega) = \sum_i q_i(\bar \omega) K(x_i, \omega).
\eeq

For this resolution function the estimator is a linear combination of the correlation function values
\beq
\bar \rho(\bar \omega) = f(\bar \omega) \sum_i q_i(\bar \omega) C(\tau_i).
\label{barrho} 
\eeq

For better approximation of $\rho(\omega)$ with the estimator $\bar \rho(\bar \omega)$ one 
needs to minimize the width of $\delta(\bar \omega, \omega)$. However, a very narrow peak might build an estimator fitting the points themselves, but not the physics (generality) they present. This means that any method of this kind should be regularized.

\begin{figure}[t!]
	\includegraphics[scale=0.4,angle=0]{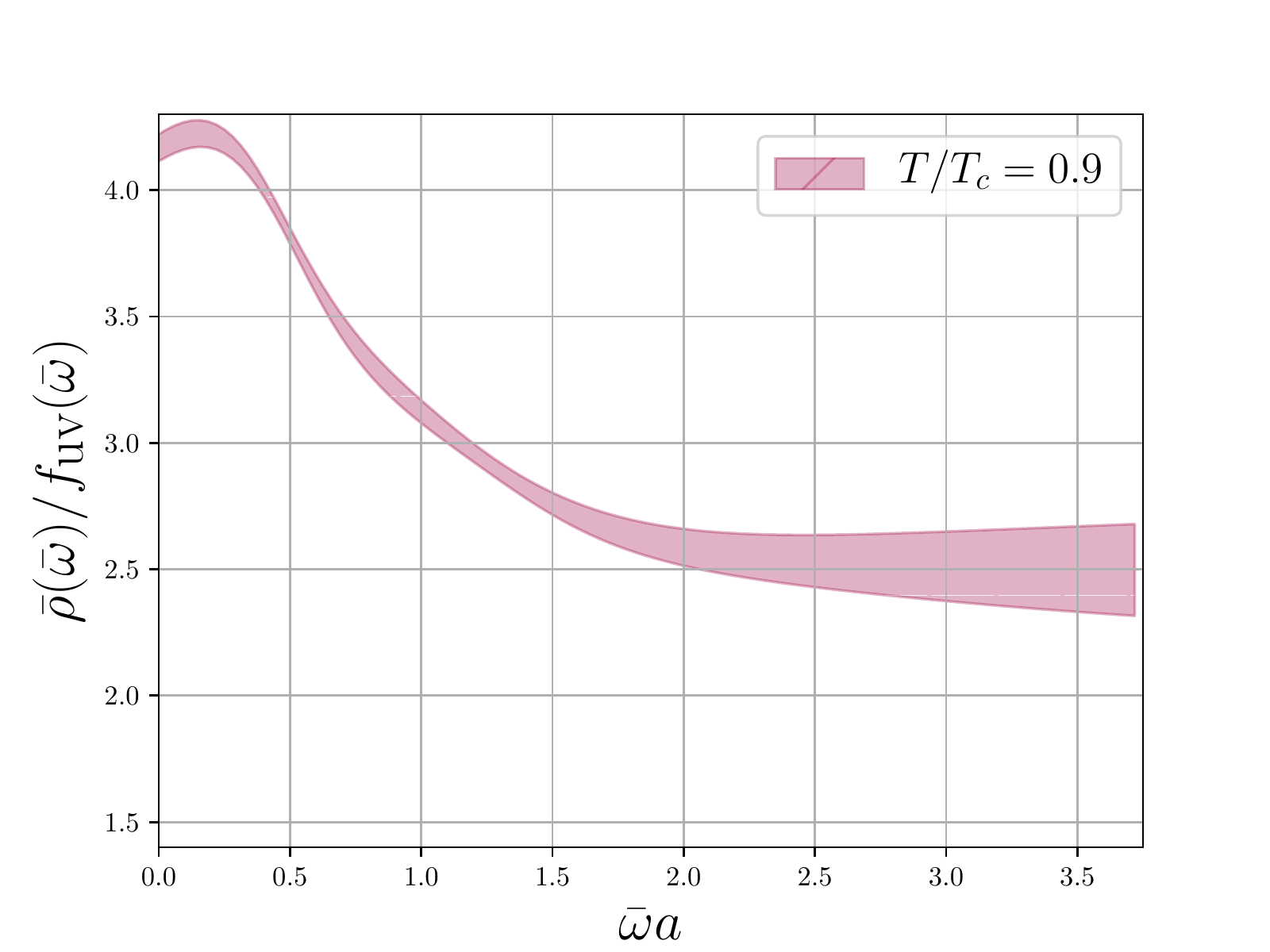}
	\includegraphics[scale=0.4,angle=0]{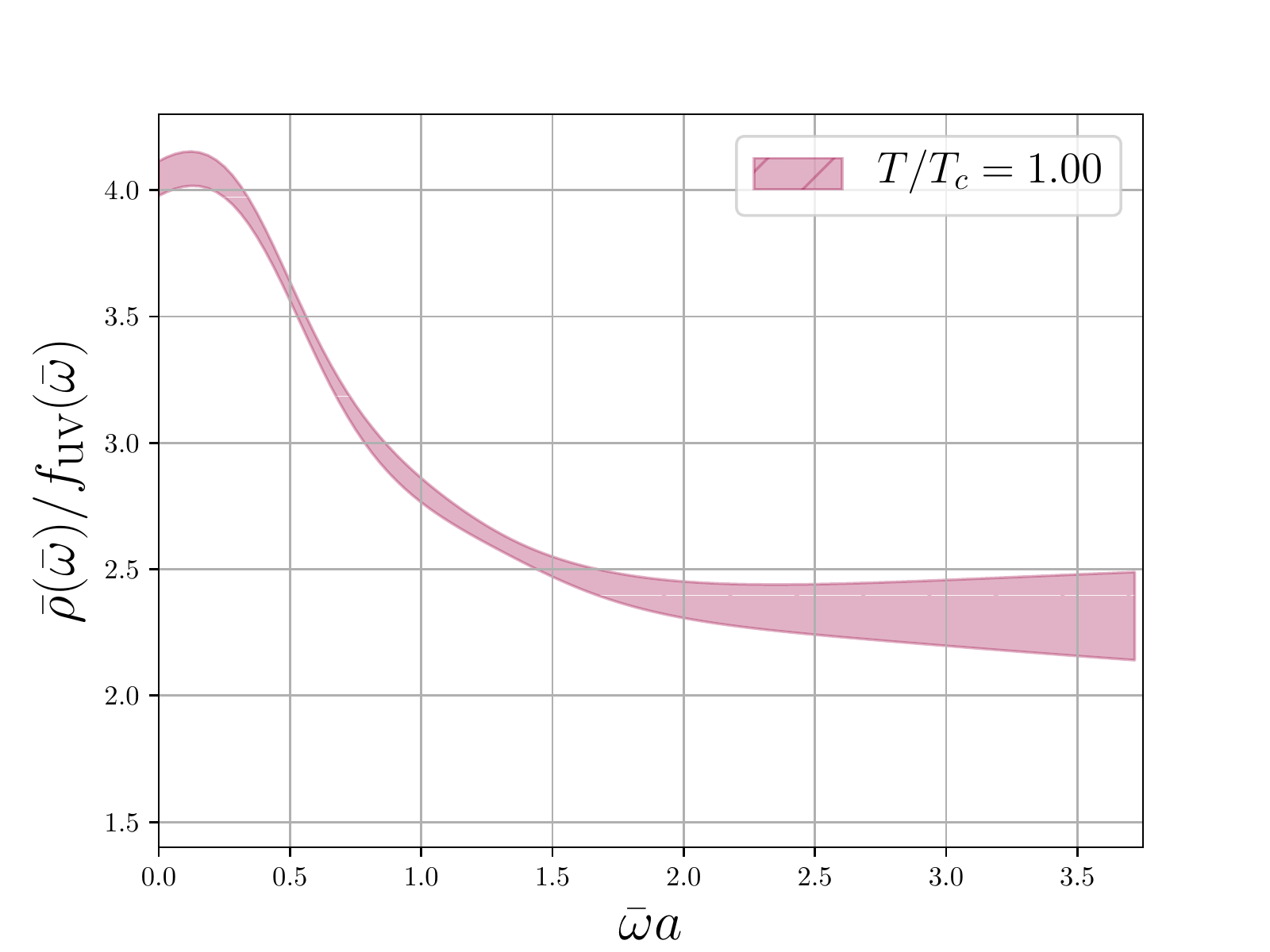} \\
	\includegraphics[scale=0.4,angle=0]{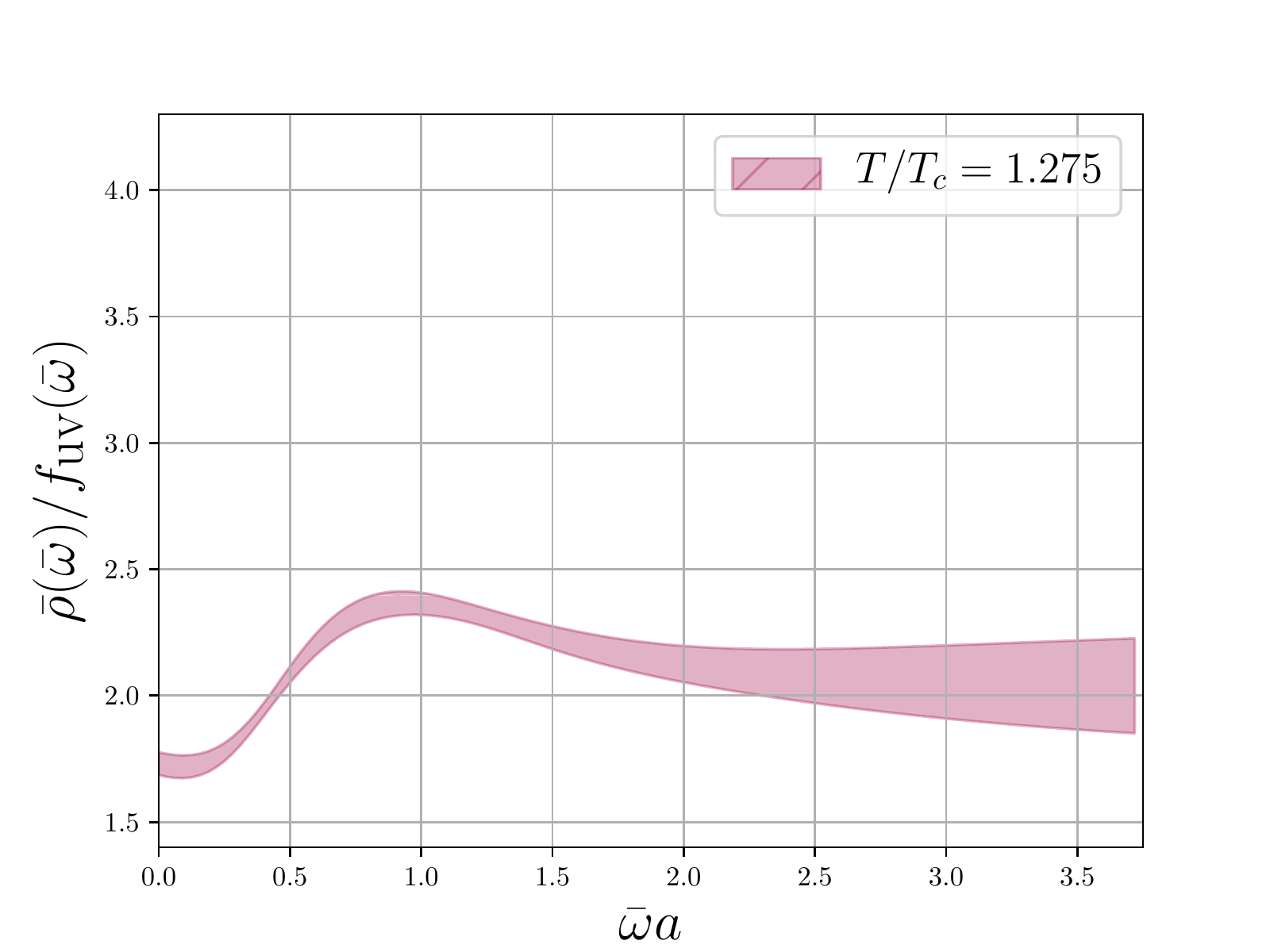}
	\includegraphics[scale=0.4,angle=0]{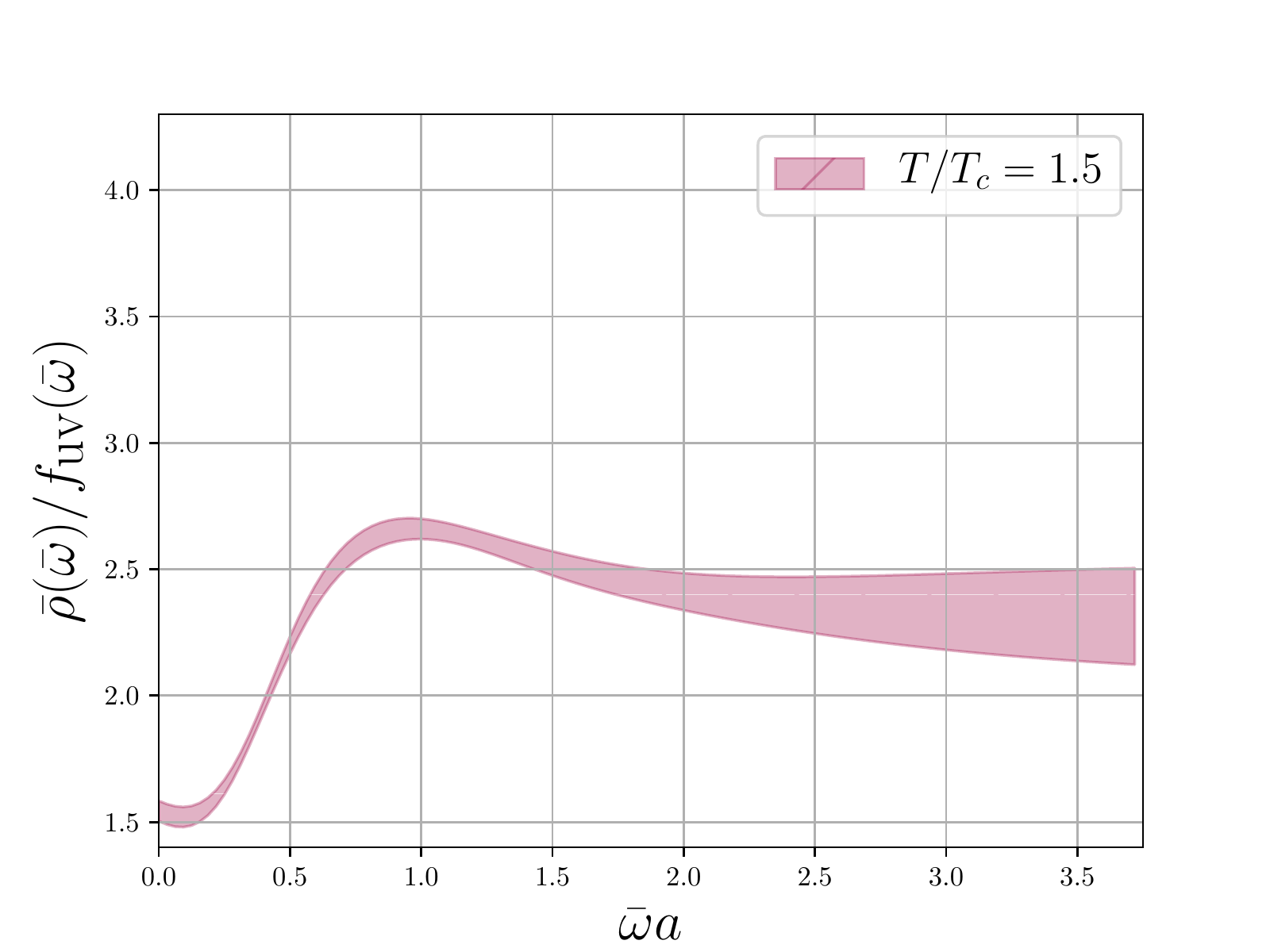}

	\caption{The ratio $\bar \rho(\bar \omega) / f_{\mbox{uv}}(\bar \omega)$ 
reconstructed within the BG method as a function of $\bar \omega a$ 
 for the temperatures: $T / T_c = 0.9,\,1.0,\,1.275,\,1.5$.}
\label{fig:bar_rho}
\end{figure}

Within the Backus-Gilbert method one minimizes the Backus-Gilbert functional $\displaystyle \mathcal{H}(\rho(\omega)) = \lambda \mathcal{A}(\rho(\omega)) + (1 - \lambda) \mathcal{B}(\rho(\omega))$. 
The term $\mathcal{A}$ represents the width of the resolution function (the second moment of distribution): $\mathcal{A} = \int_0^{\infty} d \omega \delta(\bar \omega, \omega) (\omega - \bar \omega)^2$. In principle, it could be any other function with the same meaning. The advantage of the second moment is that it is quadratic in $\omega$ and $\bar \omega$, making analytical minimization possible.
The term $\mathcal{B}(\rho(\omega)) = \mbox{Var}[\rho(\omega)]$ punishes $\rho(\omega)$ for being too dependent on the data and regularizes $\bar \rho(\bar \omega)$. In terms of the covariance matrix and $q$--functions, it reads $\displaystyle \mathcal{B}(\vec{q}) = \vec{q}^T \hat{S} \vec{q}$.

If $\lambda$ is close to $1$, the resolution function has the smallest width. However, the BG method with $\lambda \sim 1$ leads to large uncertainties. The result becomes very dependent on the data and the spectral function turns out to be noisy and unstable.
Statistical uncertainties are reduced at the expense of increasing 
the width of the resolution function through decreasing of $\lambda$. 

The minimization of $\mathcal{H}$ gives
\begin{align}
q_i(\omega) & = & \frac { \sum_j W^{-1}_{ij} (\bar \omega) R(x_j) } { \sum_{kj} R(x_k) W^{-1}_{kj} (\bar \omega) R(x_j) }, \\
W_{ij}(\bar \omega) & = & \lambda\int_{0}^{\infty} d \omega K(x_i,\omega) (\omega- \bar \omega)^2 K(x_j,\omega) + (1 - \lambda) S_{i j}, \\
R(x_i) & = & \int_0^{\infty} d \omega K(x_i,\omega). 
\label{BG_formulas}
\end{align}

In \cite{Astrakhantsev:2017nrs} it was seen that the BG method should be modified to study the shear viscosity. The estimator of the spectral function (\ref{barf})
is the convolution of the real spectral function $\rho(\omega)$ and the resolution function $\delta(\omega,\bar\omega)$, which has an ultraviolet tail. The ultraviolet behavior of $\rho(\omega)$ is $\sim \omega^4$ and it is convolved with the tail of the resolution function. 
For this reason the shear viscosity calculated within the BG method 
acquires large ultraviolet contribution which is non-physical. To get rid of this problem 
it was proposed to determine the ultraviolet tail of the spectral function and then subtract it 
from the estimator. Then the result is the convolution of only the infrared part of the 
real spectral function and the resolution function. 

To subtract the ultraviolet tail we are applied the approach proposed in \cite{Astrakhantsev:2017nrs}. The spectral function at large frequency is determined using the rescaling function 
\beq
f(x)=f_{\mbox{uv}}(x) = \alpha_s^2(x) \frac { \rho_{\mbox{lat}}(x)} {\bigl (\tanh{( x/4T)}\bigr )^2}.
\label{f1}
\eeq

We used the running coupling constant $\alpha_s$ at one-loop level with $\Lambda=237$\,MeV \cite{Capitani:1998mq}.
In the Backus-Gilbert method one reconstructs the ratio $\rho(\omega)/f_{\mbox{uv}}(\omega)$ which is divergent 
at $\omega=\Lambda$. To get rid of this divergence we assume that  $\alpha(\omega)=\alpha(1 \mbox{GeV})$ for $\omega \leqslant 1$\,GeV. The result is not sensitive to the modification of the $\alpha_s$--running since we study the large-frequency behavior of $\rho(\omega)$, which is not affected by this modification.

Finally, one has to fix the value of $\lambda$. In the BG method larger $\lambda$ leads to larger uncertainties of the calculation. 
On the other hand small $\lambda$ increases the width of the resolution function and one needs to find a compromise between these two tendencies. We found that this compromise is satisfied at $\lambda=0.01$ for our 
study of the ultraviolet properties of the spectral function. For the infrared study (see below) we used $\lambda=0.1$.

Within the BG method with (\ref{f1}) we reconstruct the ratio $\bar \rho(\bar \omega) / f_{\mbox{uv}}(\bar \omega)$. In Fig.~\ref{fig:bar_rho} we plot the reconstructed ratios for few temperatures. 
We see that at large frequencies the ratio $\bar \rho(\bar \omega) / f_{\mbox{uv}}(\bar \omega)$
reaches the plateau. The role of the $\alpha_s^2(\omega)$ factor in (\ref{f1}) should be emphasized:
if it were not for the running coupling, the ratio $\bar \rho(\bar \omega) / f_{\mbox{uv}}(\bar \omega)$ would 
not have a plateau. This means that the function $\rho_{\mbox{lat}}$ itself does not catch the essential behavior 
at large frequencies and the account of $\alpha_s^2(\omega)$ is necessary. 

\begin{figure}[t!]
	\includegraphics[scale=0.5,angle=0]{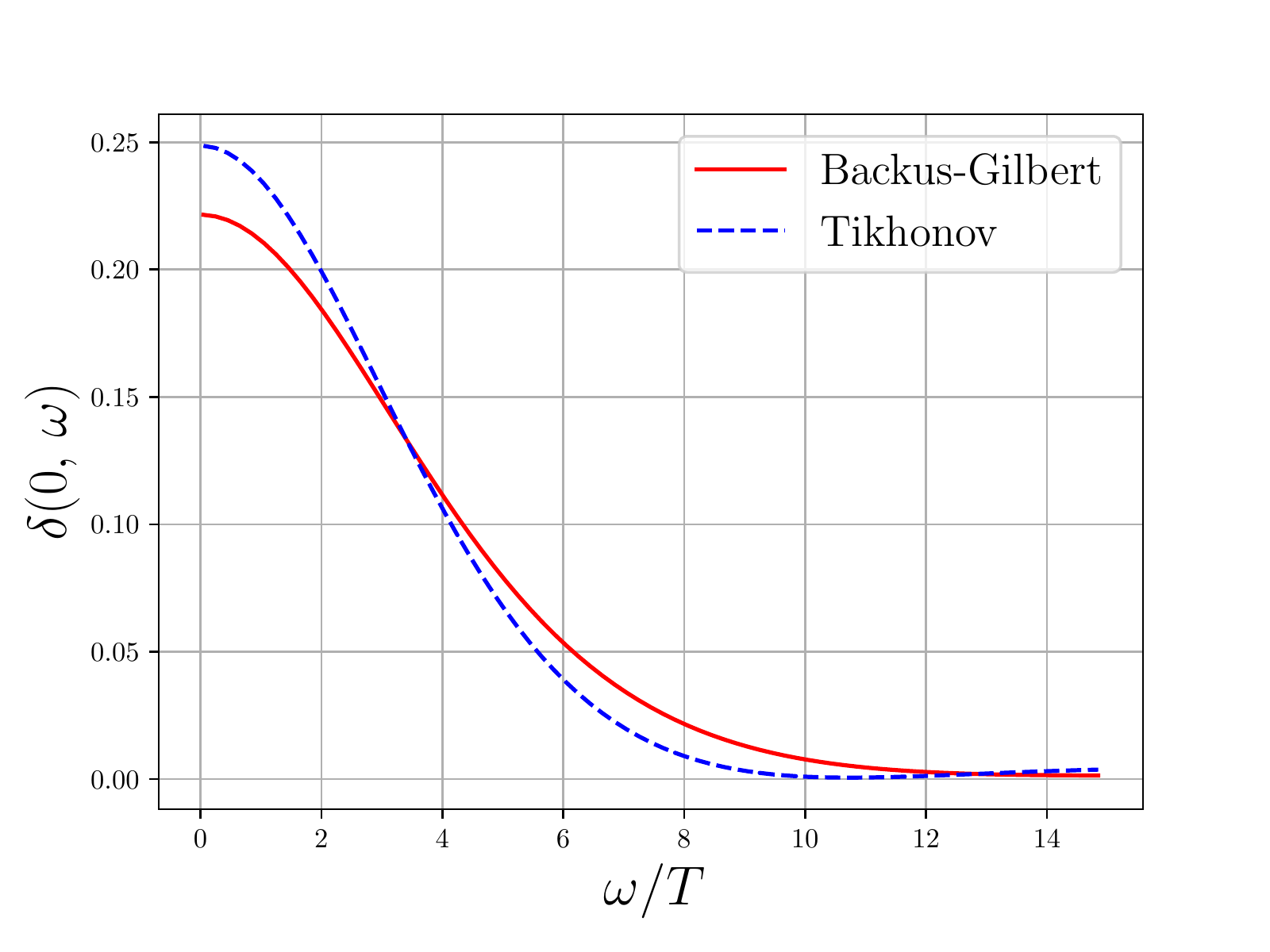}
	\caption{The resolution functions $\delta(0, \omega)$ at $T/T_c=1.05$ for the $\lambda = 0.1$ Backus-Gilbert method and for the $\gamma = 0.1$ Tikhonov regularization.}
\label{fig:deltas}
\end{figure}

We conclude that at large frequencies the 
spectral function behaves as $\rho(\omega)=A f_{\mbox{uv}}(x)$. Based on this finding we propose the following form of the ultraviolet spectral function
\begin{equation}
\rho_{uv}(\omega)=A  \alpha_s^2(\omega) \rho_{lat}(\omega) \theta(\omega-\omega_0),
\label{rho_ult}
\end{equation}
where $A$ is the value of $\bar \rho(\bar \omega) / f_{\mbox{uv}}(\bar \omega)$ on the plateau. 
In the calculation we determine the value and the uncertainties of $A$ from the plateau 
in the region $\bar \omega a \in (1.5,3)$. Another parameter of the ultraviolet tail is $\omega_0$ frequency threshold from which 
the spectral function is given by the ultraviolet form (\ref{rho_ult}). This parameter will discussed below. 

Having determined the ultraviolet behavior of the spectral function, we proceed 
to the calculation of the bulk viscosity. 
In order to calculate the $\zeta$, we found the estimator for $\rho(\omega) / \omega$ at $\bar \omega = 0$. 
We calculate it using the BG method with $f(x)=x$. 

The resolution function for the temperature $T/T_c=1.05$ is shown in Fig.~\ref{fig:deltas} (solid line). The resolution functions at the other temperatures are close to that for the $T/T_c=1.05$.  It is seen that the width of the resolution function is $\sim 4 T$. Thus the spectral function $\rho(\omega) / \omega$ is averaged over the region $\sim 4 T$. 

With the resolution function we calculate the estimator for $\rho(\omega) / \omega$. 
Then we subtract the ultraviolet contribution given by the convolution of the 
spectral function (\ref{rho_ult}) and the resolution function. The threshold parameter $\omega_0$ can not be determined within the BG analysis. To account its uncertainty, we vary $\omega_0$ within the region $\omega_0/T_c \in (5,10)$ $\Big(\omega_0\sim (1.4, 2.8)\text{ GeV}\Big)$.
We believe that this region is sufficiently safe to estimate the uncertainty due to the unknown value of $\omega_0$.

\begin{figure}[h!]
	\begin{center}
		\includegraphics[scale = 0.5]{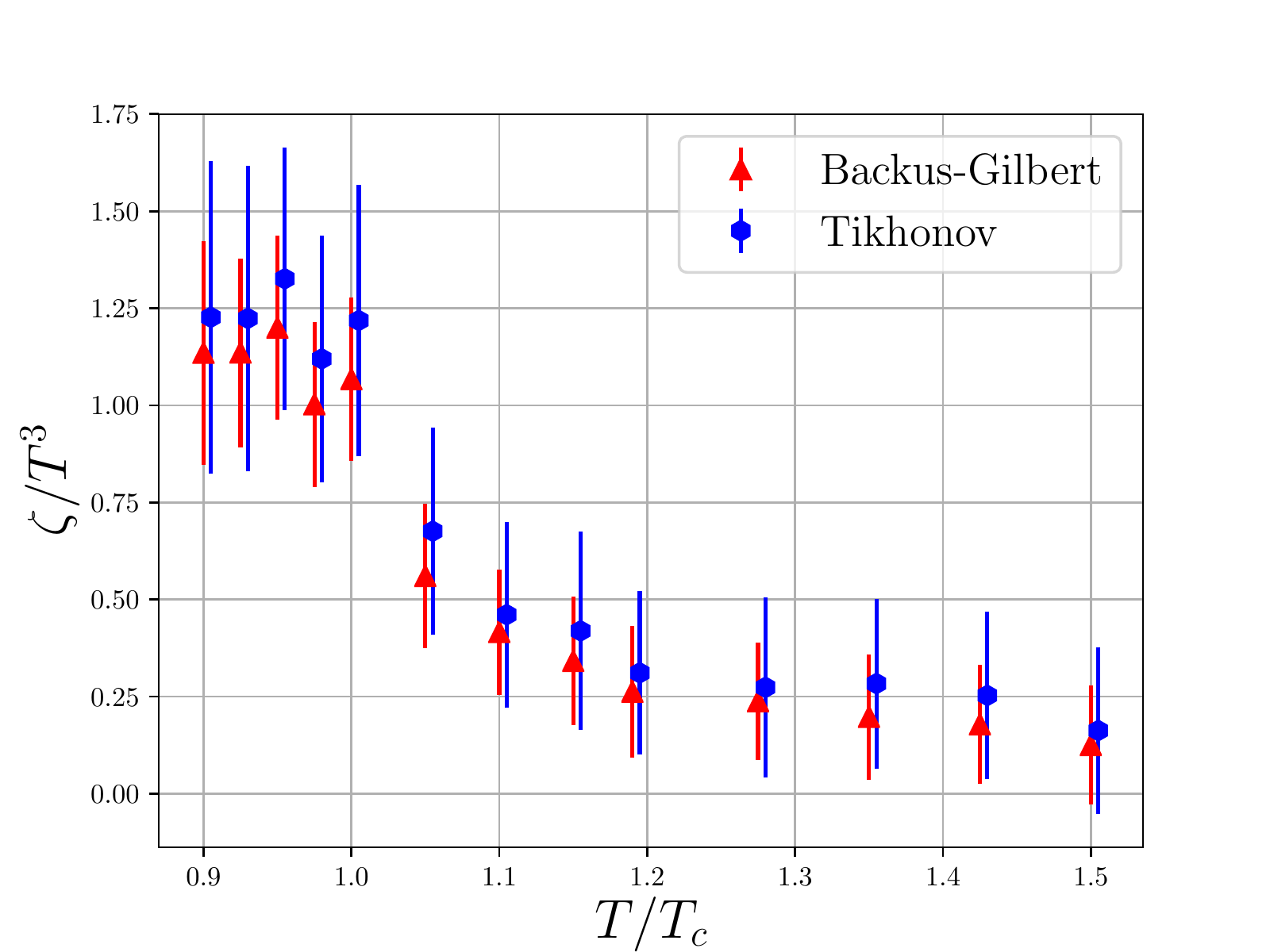}
        \includegraphics[scale = 0.5]{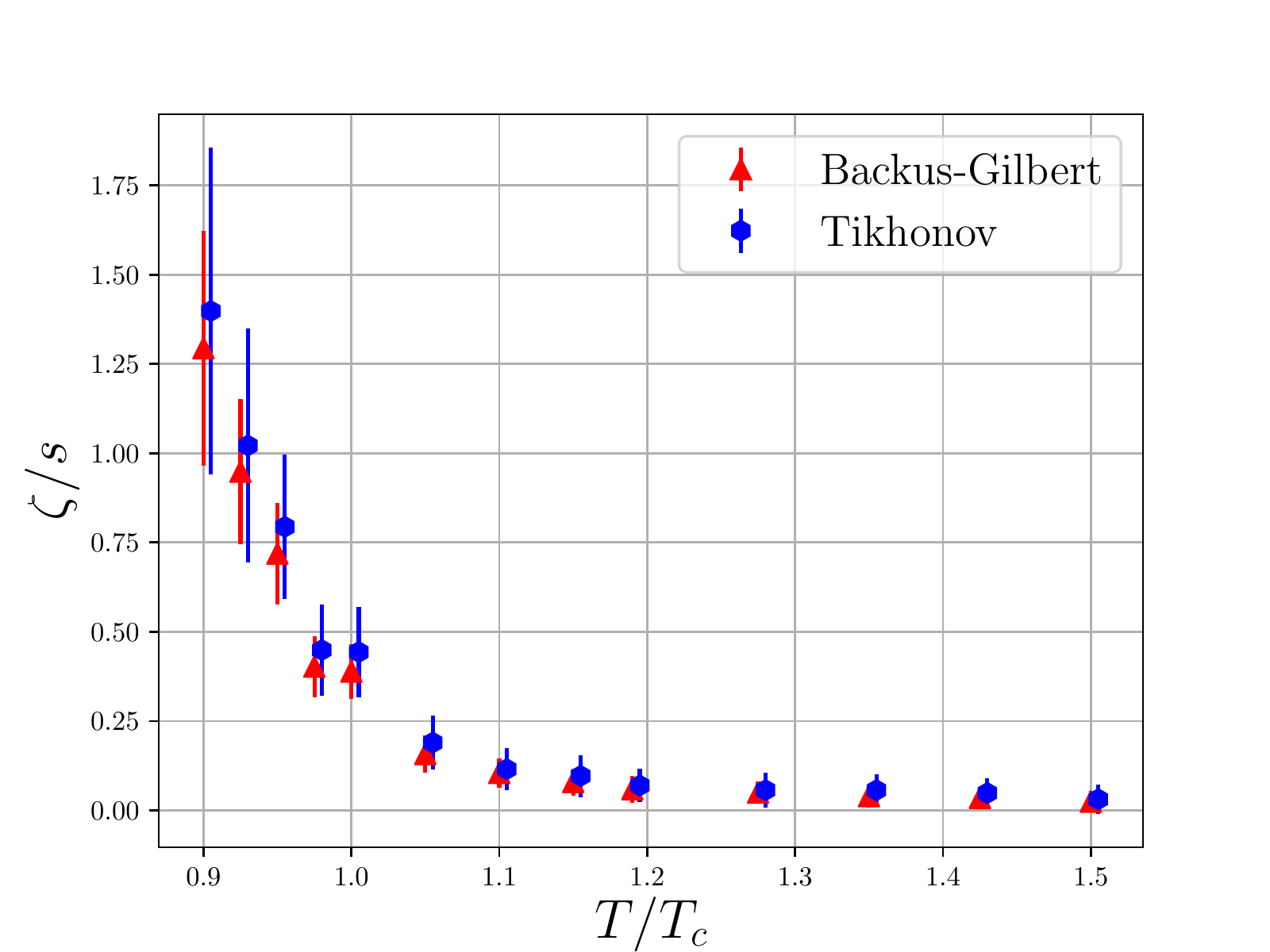}
		\caption{The ratios $\zeta / T^3$ and $\zeta / s$ 
calculated within the Backus-Gilbert and the Tikhonov regularization methods        
 as functions of temperature.}
		\label{fig:final_plots}
	\end{center}
\end{figure}

Our results for the ratios $\zeta/T^3$ and $\zeta/s$ are shown in Fig.~\ref{fig:final_plots}. 
The uncertainties shown in Fig.~\ref{fig:final_plots} are due to statistical errors and the 
uncertainties in the $A$ and $\omega_0$.

\subsection*{Calculation of the bulk viscosity using the Tikhonov regularization}
Finally we consider another approach to the bulk viscosity estimation, called the Tikhonov regularization (TR)\footnote{The TR method was recently applied to study of the metal-insulator phase transition in Hubbard model~\cite{Ulybyshev:2017szp}.} \cite{Tikhonov:1963}. 
The TR method allows to make the resolution function narrower
as compared to the BG method.
In the calculation with the TR we follow the formulas (\ref{BG_formulas}). 
The difference of the TR as compared to the BG method is in the regularization 
of the $W_{ij}$ matrix. In the BG method one adds the covariance matrix $\lambda S_{ij}$
to the matrix $(1 - \lambda) W_{ij}$. In the TR method the matrix $W_{ij}$ is regularized as follows.
Let us consider the SVD decomposition of the $W^{-1}$: $W^{-1} = V D U^{T}$, where $D = \mbox{diag}\left(\sigma_1^{-1}, \sigma_2^{-1}, \ldots, \sigma_n^{-1}\right)$. We substitute the matrix $D$ by the matrix $\tilde D = \mbox{diag}\left((\sigma_1 + \gamma)^{-1}, (\sigma_2 + \gamma)^{-1}, \ldots, (\sigma_n + \gamma)^{-1}\right)$, where the $\gamma$ is the regularization parameter of the TR method. The TR method thus smoothly cuts-off small singular values $\sigma$, making the results more stable.

The application of the TR method is the same as the application of the BG, but with the $W_{ij}$ regularized in the other way. We first choose the regularization 
parameter $\gamma = 0.1$ which is a compromise between the width of the resolution function 
and the uncertainty of the calculation. The resolution function in this case is shown in Fig.~\ref{fig:deltas} and it has the width $\sim 3 T$ which is smaller than in the BG method. We then calculate the estimator with $f(x)=x$ and subtract the ultraviolet tail. The ultraviolet spectral function was taken in the form (\ref{rho_ult})
with $A$ and $\omega_0$ from the BG method. In Fig.~\ref{fig:final_plots} we plot our results.

\section*{\large Discussion and conclusion}

In this paper the temperature dependence of the bulk viscosity in gluodynamics is studied within lattice simulation. 
To carry out this study we measured the correlation functions of the
energy-momentum tensor trace for a set of temperatures in the range $T/T_c \in (0.9, 1.5)$.
To extract the bulk viscosity from the correlation function we applied the
Backus-Gilbert method and the Tikhonov regularization method. The results obtained 
within both approaches are shown in Fig.~\ref{fig:final_plots}. 
We also studied the finite volume effects and found that
they are small for all the temperatures except the $T=T_c$. Finite volume effects for $T=T_c$ might be important. 

Let us now consider the results obtained in this paper. 
From Fig.~\ref{fig:final_plots} we see that the ratio $\zeta/s$ is 
small in the region $T/T_c \geqslant 1.1-1.2$ and in the vicinity of the transition $T/T_c \leqslant 1.1-1.2$ it quickly rises. This behavior is 
in agreement with a lot of phenomenological studies (see, for instance, \cite{Kharzeev:2007wb, Saha:2017xjq, Marty:2013ita}).

Below the critical temperature $\zeta/s$ continues to rise. It
is in disagreement with some phenomenological studies of QCD~\cite{Berrehrah:2016vzw,Singha:2017jmq}. 
This discrepancy might be explained as follows. In our study we
convolute the spectral function with the resolution function which has 
the width $\sim (3-4) T_c$. Below the critical temperature there is the scalar glueball contribution to the spectral function which might alter our result.
One might also find another explanation of this discrepancy. In is known that the confinement/deconfinement phase transition in the $SU(3)$--gluodynamics is of the first order, while this transition in QCD is a crossover. The rise of the $\zeta/s$ below the critical point might be assigned to the rapid decrease of the entropy density $s$ below the transition in the gluodynamics. 
Better understanding of this discrepancy requires additional study of the bulk viscosity and the spectral function of the energy-momentum tensor both from the lattice side and in these phenomenological models.

\begin{figure}[h!]
	\begin{center}
		\includegraphics[scale = 0.5]{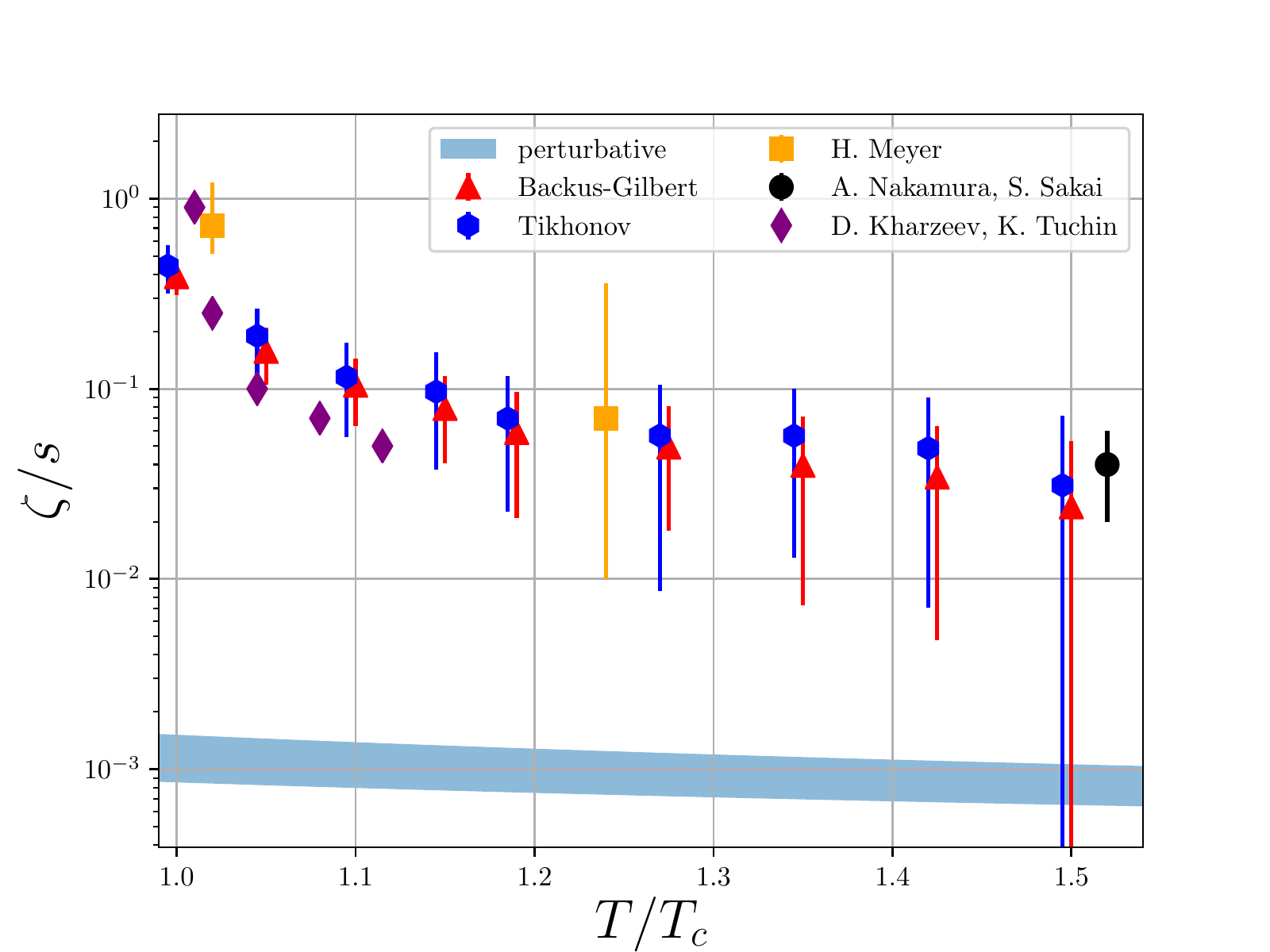}
		\caption{The ratio $\zeta/s$ calculated in this paper and other studies: the results 
obtained in this paper within the Backus-Gilbert method and the Tikhonov regularization, the lattice results obtained in \cite{Nakamura:2004sy} and \cite{Meyer:2007dy}, perturbative results obtained in \cite{Arnold:2006fz} and the results of \cite{Kharzeev:2007wb}.
}
		\label{fig:data_compare}
	\end{center}
\end{figure}

In order to compare our results with the results of other approaches, 
in Fig.~\ref{fig:data_compare} we plot the ratios $\zeta/s$ for $T/T_c \geqslant 1$ calculated in our paper 
and in other studies. In particular, the blue circles and the red triangles represent the results 
obtained in this paper within the Backus-Gilbert method and the Tikhonov regularization correspondingly. 

The black circles and the yellow squares represent lattice results obtained in \cite{Nakamura:2004sy} and \cite{Meyer:2007dy} 
correspondingly. It is seen that our results are in agreement with the previous lattice studies of the bulk viscosity. 

The blue band represents the perturbative results obtained in \cite{Arnold:2006fz}. 
The uncertainty in this band is due to the variation of the scale in the region $\mu \in (2\pi T,4 \pi T)$.
It is seen that out results dramatically disagree with the perturbative results, 
what once again confirms that QGP is a strongly correlated system. 

The results of \cite{Kharzeev:2007wb} are represented with the violet diamonds. It is interesting to notice that the rise of $\zeta/s$ in \cite{Kharzeev:2007wb} starts at $T/T_c \sim 1.1$ which agrees with our results.

\begin{figure}[h!]
	\begin{center}
		\includegraphics[scale = 0.5]{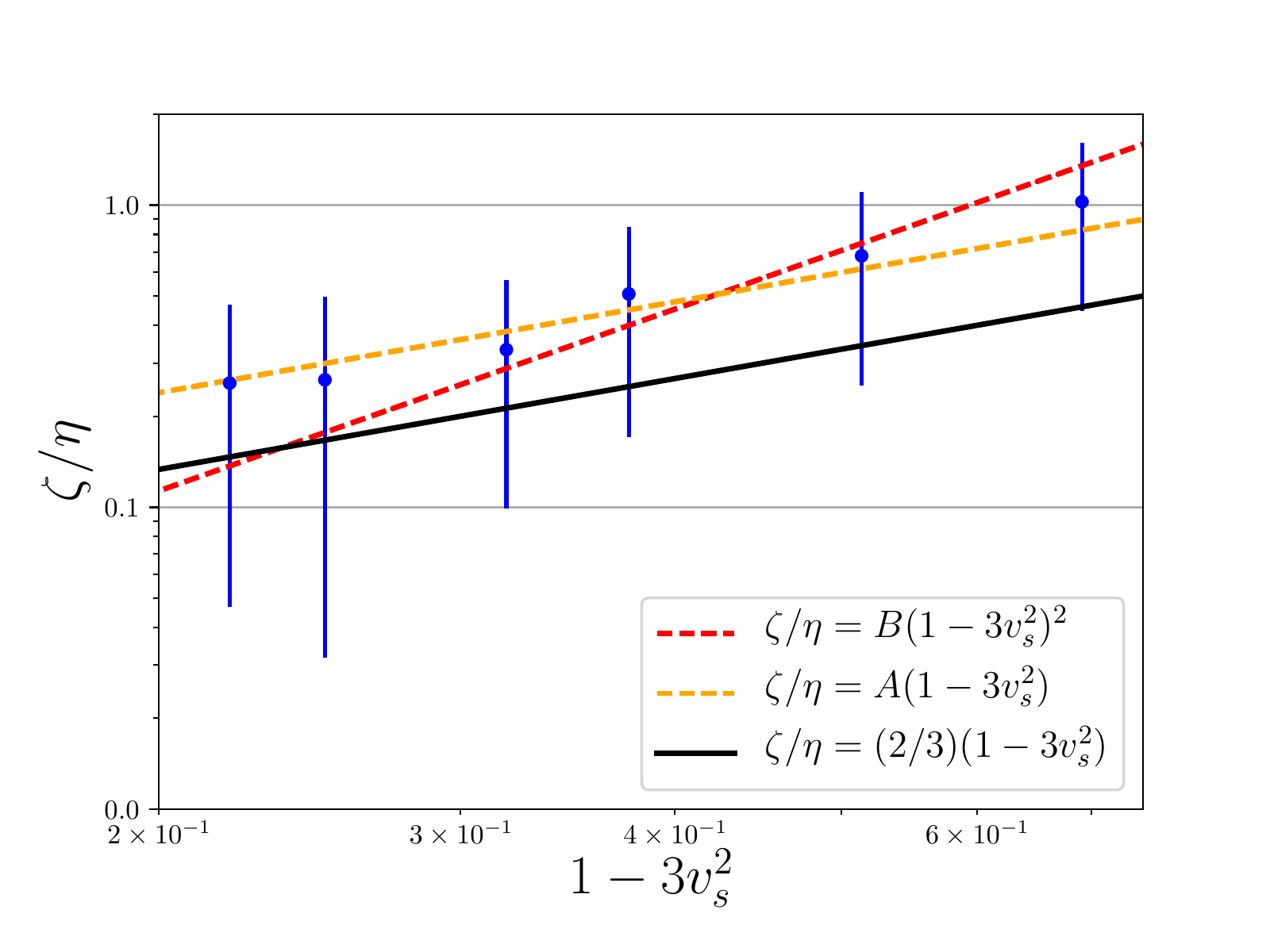}
		\caption{The ratio $\zeta/\eta$ as the function the velocity of sound.  The orange line corresponds to the best linear fit, the red line shows the best quadratic fit. The black line represents the lower bound of the ratio $\zeta / \eta$ imposed by the AdS/CFT calculations.}
		\label{fig:fits}
	\end{center}
\end{figure}

Finally let us consider the following question. The perturbative calculations of the $\zeta$ 
revealed the following relation between the bulk viscosity, the shear viscosity $\eta$ and 
the speed of sound $v_s$: $\zeta/\eta \propto (1-3 v_s^2)^2$. On the other hand similar ratio 
in the AdS/CFT is predicted to be: $\zeta/\eta \propto (1-3 v_s^2)$~\cite{Benincasa:2005iv, Buchel:2007mf}. 
In addition it was argued that there is an inequality $\zeta/\eta \geqslant 2/3 (1-3 v_s^2)$ which is valid for QGP~\cite{Buchel:2007mf}. 

To check these assumptions in Fig.~\ref{fig:fits} we plot our results obtained within the Tikhonov regularization\footnote{One can plot the similar picture for the Backus-Gilbert results.}
in the region $T/T_c \in (1.05, 1.425)$ as a function of the speed of sound in gluodynamics 
calculated in \cite{Boyd:1996bx}. The viscosity $\eta$ was taken from our paper \cite{Astrakhantsev:2017nrs}.
If the temperature at which $\zeta$ is calculated is not present in \cite{Astrakhantsev:2017nrs} 
we take the average of the closest points. We 
fit our data with the linear and the quadratic fits. We also plot the line $\zeta/\eta = 2/3 (1-3 v_s^2)$. 
Unfortunately the uncertainty of the calculation is rather large and one can not distinguish between these two hypotheses.

\acknowledgments

V.\,V.\,B. acknowledges the support from the BASIS foundation.
N.\,Yu.\,A. acknowledges the support from the BASIS foundation and FAIR-Russia Research Center.
The work of  A.\,Yu.\,K was supported by FAIR-Russia Research Center and RFBR grants  18-32-00071.
This work has been carried out using computing resources of the federal 
collective usage center Complex for Simulation and Data Processing for 
Mega-science Facilities at NRC ``Kurchatov Institute'',~\url{http://ckp.nrcki.ru/}. In addition, we used the supercomputer of the
Institute for Theoretical and Experimental Physics (ITEP).

\end{document}